\documentclass[conference]{IEEEtran}
\IEEEoverridecommandlockouts
\usepackage{comment,color,cite,url}
\usepackage{subfigure,times,graphicx,epsfig}
\usepackage{amsmath}
\usepackage{amssymb}
\usepackage{mathtools}
\usepackage{algorithm}
\usepackage{algpseudocode}

\def\BibTeX{{\rm B\kern-.05em{\sc i\kern-.025em b}\kern-.08em
    T\kern-.1667em\lower.7ex\hbox{E}\kern-.125emX}}

\DeclareMathAlphabet{\mathsfbf}{OT1}{cmss}{sbc}{n}

\newcommand{\EE}{\mathbb{E}} 

\newcommand{\ee}{{\rm e}}
\newcommand{\dd}{{\rm\,d}} 

\newcommand{\av}{{\bf a}}
\newcommand{\bv}{{\bf b}}
\newcommand{\cv}{{\bf c}}

\newcommand{\hv}{{\bf h}}

\newcommand{\uv}{{\bf u}}

\newcommand{\yv}{{\bf y}}

\newcommand{\Cc}{{\cal C}}

\newcommand{\Hc}{{\cal H}}

\newcommand{\Nc}{{\cal N}}

\newcommand{\Sc}{{\cal S}}

\newcommand{\Wc}{{\cal W}}

\newcommand{\Xc}{{\cal X}}
\newcommand{\Yc}{{\cal Y}}

\newcommand{\rhov}{\boldsymbol{\rho}}

\def\ben{\begin{enumerate}}
\def\beq{\begin{equation}}
\def\beqa{\begin{eqnarray}}
\def\bit{\begin{itemize}}
\def\een{\end{enumerate}}
\def\eeq{\end{equation}}
\def\eeqa{\end{eqnarray}}
\def\eit{\end{itemize}}
\def\bef{\begin{figure}}
\def\eef{\end{figure}}
\def\bec{\begin{center}}
\def\eec{\end{center}}

\begin{document}
\title{Uplink soft handover for LEO constellations: how strong the inter-satellite link should be}

 \author{
      \IEEEauthorblockN{1\textsuperscript{st} Houcem Ben Salem}
\IEEEauthorblockA{\textit{CNR-IEIIT} \\
Torino, Italy \\
houcem.bensalem@ieiit.cnr.it}
\and
\IEEEauthorblockN{2\textsuperscript{nd} Alberto Tarable}
\IEEEauthorblockA{\textit{CNR-IEIIT} \\
Torino, Italy \\
alberto.tarable@cnr.it}
\and
\IEEEauthorblockN{3\textsuperscript{rd} Alessandro Nordio}
\IEEEauthorblockA{\textit{CNR-IEIIT} \\
Torino, Italy \\
alessandro.nordio@cnr.it}
\and
\IEEEauthorblockN{4\textsuperscript{th} Behrooz Makki}
\IEEEauthorblockA{\textit{ Ericsson EAB} \\
G\"oteborg, Sweden \\
behrooz.makki@ericsson.com}    
\thanks{}}

\maketitle

  \begin{abstract}
    We consider a constellation of low-earth-orbit (LEO) satellites connected to a
    handheld device on the ground.  Due to the very large orbital
    speed, an effective handover strategy becomes of paramount
    importance. In particular, we study the benefits of soft handover
    in the uplink from the physical-layer point of view.  We give a
    realistic model for both the ground-to-satellite and the
    inter-satellite links,  following the 3GPP channel model for the
    former.  We suppose that, during handover from a serving satellite to a target satellite,  one of the two satellites forwards the
    received signal from the ground user to the other,
    thus acting as a relay.  We quantify through simulations the loss of  hard handover, compared to soft handover.  For the latter,  we test both amplify-and-forward (AF) and
    decode-and-forward (DF) relaying techniques and verify that,  at least in the simulated conditions,  DF does not repay,  in terms of block error rate (BLER), the increase of complexity with respect to AF.  Also, we study the effect of the LEO constellation size on the network BLER.   Finally,  we show that, with soft handover,  the impact of misalignment on the inter-satellite link is severe,  especially at optical frequencies. 
\end{abstract}
\begin{IEEEkeywords}
 Non-Terrestrial Networks,  LEO constellations,  soft handover,  Inter-Satellite Link
\end{IEEEkeywords}



\section{Introduction}\label{sec:introduction}

The integration of terrestrial and non-terrestrial networks (NTNs) is
foreseen as one of the key factors that will enable ubiquitous
connectivity in future 6G networks
\cite{Gustavsson21,rajatheva2021scoring}.  NTNs can also play a
fundamental role in ensuring connectivity in cases where the
terrestrial infrastructure is missing or indequate to support the
requested traffic, such as in rural areas or in disaster scenarios.
In particular, low-earth orbit (LEO) NTNs have now become an important
research subject, also fostered by the surge of privately-owned LEO
constellations put into operations by several companies.

LEO NTNs offer smaller round-trip times (up to 30 ms) compared to geostationary NTNs, but they present a number of challenges.  Since a LEO satellite provides a limited
coverage on the earth surface,   several tens of 
satellites need to be employed in the same LEO, in order to ensure
uninterrupted connectivity with static ground users (GU). 
Even more importantly,  LEO satellites have a very short orbital period.  As an
example, a satellite having height 550\,km above the earth surface performs a revolution
in about 1.6 hours.  
Thus,  a LEO satellite remains in visibility for a limited amount of time
(a few minutes),  so that there is need for frequent handover, even for a
static GU~\cite{Juan22}.  In this regard, we can
distinguish two different system-level choices.  In the first method,  each
satellite steers its antenna array so as to point always to the same ground
area,  therefore the cells are earth-fixed.  In the second method,  the satellite
array points to a fixed direction, so that the cells are earth-moving.
In the first case, handover between different satellites has to be
performed every few minutes.  In the second case, handover events are
much more frequent, once every few seconds \cite{Juan22}.  
In both scenarios,  it is clear that handling effectively the handover is crucial to ensure a high-level quality of service.

In this paper,  we consider the uplink (UL) of a satellite communication
system where a GU initially communicates with a serving satellite
S. In an ideal hard-handover scenario, the GU instantaneously switches
to a target satellite  T, when its elevation on the horizon becomes
greater than that of S (for a more realistic protocol, see, 
e.g.,  \cite{Juan22}). Instead, in soft handover the GU signal is
received by both S and T for a short period of time. Then,  the lower-elevation satellite (which is first T, then S) acts as a relay,  by  forwarding
its received signal to the higher-elevation one through the inter-satellite link (ISL).  
In the following,  we are particularly interested in
understanding the role of the ISL to help the handover.
We study the block error rate (BLER) when 
amplify-and-forward (AF) and decode-and-forward (DF) relaying
techniques are applied.  We consider realistic 
models for ground-to-satellite (G2S) and ISL channels,
following the recent 3GPP model for the former. Also, we assess the advantage of soft handover with respect to hard handover,  and we study the
effect of the size of the satellite constellation and of the misalignment in the ISL on the performance of
soft-handover-based LEO systems. 

A paper dealing with soft handover in the UL is \cite{Barros13}, which
considers multipacket transmission on a highly time-dispersive
channel.  However, the ISL is assumed ideal, and the satellite
performs linear minimum-mean square error filtering on the signals
received from several GUs.  A paper that performs multi-satellite
reception (without reference to handover) is the recent
\cite{Farbod23}. In that context, the channel model defined in
\cite{3GPPTR38811,ITURRECP68111} is used and imperfect channel state
information is taken into account.  Performance is measured in terms
of achievable capacity.  However, also in that case, the ISL is
assumed ideal.

Differently from previous works, we precisely address the role of the
ISL in the performance of soft handover.  We perform evaluation of
BLER by simulating a realistic environment in which the time evolution
of satellite elevations is obtained by geometrical considerations, the
G2S link is based on the 3GPP standard channel
model defined in \cite{3GPPTR38811,ITURRECP68111}, and the ISL channel
model is based on \cite{Akyildiz21}. \footnote{For the ISL,  a very recent paper \cite{Guven23} proposes a multi-state  channel model, which will be used in the extensions of the present work.}  

We will address the following questions:
\begin{itemize}
\item What is the advantage of soft handover, compared to hard
  handover?

\item What is the best strategy the relaying satellite can
  adopt? To answer this question, we will compare the performance of AF
  and DF relaying techniques.
  \item What is the impact of the constellation size?
  \item How severe is the impact of misalignment for an optical ISL? In this respect,  is an ISL in the THz band more robust?
\end{itemize}

In the following, we partially answer such questions,  by concluding that,  with a proper choice of the soft-handover scheme, and with a suitable design of the ISL,  the performance gain of soft handover over hard handover justifies the additional complexity related to its implementation.

\section{System description\label{sec:model} }

\subsection{Geometrical Model\label{sec:geo_model}}

\begin{figure}[t]
  \centering
\includegraphics[width=0.7\columnwidth]{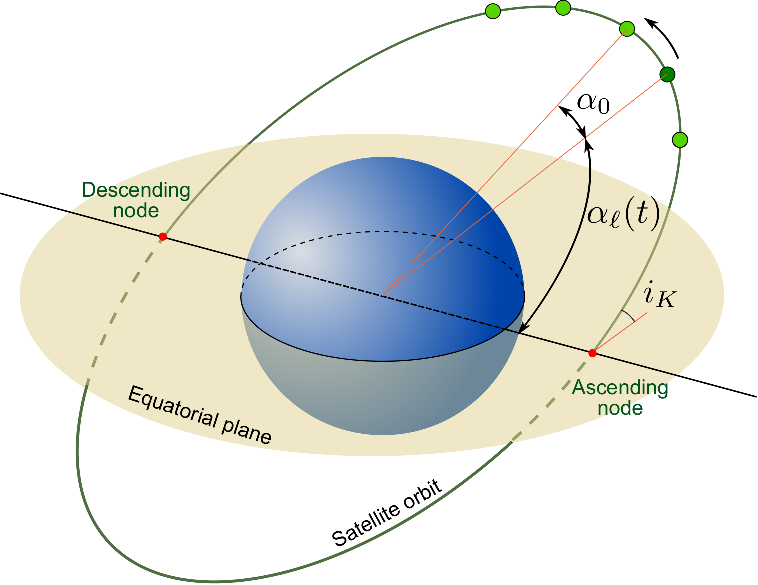}
\caption{Swarm of $M$ satellites deployed on a circular orbit with
  inclination $i_{\rm K}$ with respect to the earth equatorial
  plane. Adjacent satellites are angularly spaced by $\alpha_0$
  radiants.}
\label{fig:sat}
\end{figure}

\begin{figure}[t]
  \centering
\includegraphics[width=0.6\columnwidth]{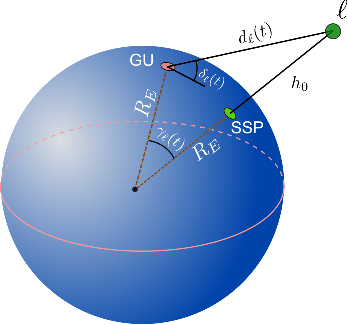}
\caption{Geometric model of satellite $\ell$ orbiting the earth at height
  $h_0$. The satellite elevation angle, as observed by a GU, is denoted by $\delta_{\ell}(t)$.}
\label{fig:sat2}
 \end{figure}

We consider a set of $M$ satellites deployed on a circular orbit, with
height on the ground $h_0$ and inclination $i_{\rm K}$ with respect to
the equatorial plane. Satellites are angularly spaced on the orbit by
$\alpha_0 = 2\pi/M$ radiants, as depicted in Figure \ref{fig:sat}.
According to Kepler's laws, the angular speed of each satellite is
given by \cite{richharia2010satellite} \beq \omega =
\sqrt{\frac{\mu}{(R_{\rm E} + h_0)^3}}\quad \mbox{[rad/s]}, \eeq where
$R_{\rm E} = 6371$\,km is the earth radius and
$\mu = 3.986 \times 10^5$\,km$^3$s$^{-2}$ is the gravitational
constant. Let us denote by $\ell$ either the serving
satellite,  $\ell = S$, or the target satellite,  $\ell = T$.  Let
$\alpha_{\ell}(t) = \omega t+\alpha_{\ell}(0)$, $\ell\in\{{\rm S},{\rm T}\}$, be the
angular position of satellite $\ell$ at time $t$.  Without loss of generality,  we can set $\alpha_{\rm S}(0)=0$
and $\alpha_{\rm T}(0)=\alpha_0$, i.e., S and T are adjacent satellites on the orbit.
Also,  let $\lambda_{\ell}(t)$ ad $\phi_{\ell}(t)$ be the latitude and longitude
of the sub-satellite point (SSP)\footnote{The SSP of satellite
  $\ell$ is the point on the earth surface lying on the straight line
  connecting $\ell$ to the earth center.} of $\ell$ at time $t$.  By geometric
considerations, we obtain
\begin{eqnarray}
\lambda_{\ell}(t) &=& \arcsin \left( \sin i_K \sin \alpha_{\ell}(t) \right) \\
\phi_{\ell}(t) & = & \mathrm{mod}\left(\widetilde{\phi}_{\ell}(t) - \omega_{\rm E} t + \pi, 2\pi \right) - \pi, \label{eq:phis}
\end{eqnarray}
where $\omega_{\rm E}$ is the earth angular speed, 
\begin{equation}
\widetilde{\phi}_{\ell}(t) = \mathrm{atan2} (\cos i_{\rm K} \sin \alpha_{\ell}(t), \cos
\alpha_{\ell}(t)), 
\end{equation}
and $\mathrm{atan2}(\cdot, \cdot)$ is the 4-quadrant inverse tangent
function.  With
reference to Fig.~\ref{fig:sat2}, let $\lambda_{\rm GU}$ and $\phi_{\rm
  GU}$ be latitude and longitude of the GU, supposed fixed in a single
satellite pass\footnote{Since the satellite is in sight for a few
minutes, a pedestrian (5 km/h) or even a vehicular (50 km/h) mobile
speed can meet this hypothesis.}.  Using the spherical law of cosines,
the central angle between the GU and the SSP of satellite $\ell$ is given by
\cite{Kells40}
\begin{equation} \gamma_{\ell}(t) = \arccos \left( \sin \lambda_{\ell}(t)
\sin \lambda_{\rm GU} + \cos \lambda_{\ell}(t) \cos \lambda_{\rm GU}
\cos \Delta \phi_{\ell}(t) \right)
\end{equation}
where $\Delta \phi_{\ell}(t) = \phi_{\rm GU}-\phi_{\ell}(t)$.  The elevation of
the satellite $\ell$ at time $t$, as observed by the GU, can be found using
the law of sines \cite{richharia2010satellite}, yielding
\beq
\delta_{\ell}(t) = \arctan \frac{\cos \gamma_{\ell}(t) - \frac{R_{\rm E}}{R_{\rm E}
    + h_0}}{\sin \gamma_{\ell}(t)}\,.
\eeq
We define $\delta_{\min}$ as the
minimum elevation that allows a successful communication between the
GU and the satellite.  Finally, considering Fig. \ref{fig:sat2}, the
slant distance of the satellite $\ell$ from the GU can be obtained by the law
of cosines \cite{3GPPTR38811} as
\beq d_{\ell}(t) = \sqrt{R_{\rm E}^2 \sin^2
  \delta_{\ell}(t) + h_0^2 + 2 h_0 R_{\rm E} } - R_{\rm E} \sin \delta_{\ell}(t)
\eeq

\subsection{Ground-to-satellite Channel Model}
We assume the narrow-band G2S channel model presented in
\cite{3GPPTR38811,ITURRECP68111}.  It is based on a Semi-Markov chain,
where there are two states, good (G) and bad (B). State G represents a
condition in which the effect of obstructions and fading is relatively
low, while in state B shadowing and fading have a more severe impact
on the received signal-to-noise ratio (SNR). The evolution of the
channel conditions corresponds to an alternate sequence of states G
and B, where the duration of each state occurrence is a
random variable independent of all the others. In both states,   the
duration has a lognormal probability density function (PDF), with
different parameters for state-G and state-B occurrences.

Within a given state occurrence, the channel coefficients obey a Loo
distribution, which is the superposition of a lognormal shadowing
(modeling slow channel variations) and a Rayleigh fading.  Under this
model, a channel coefficient is a complex random variable with
independent magnitude and phase. The phase is uniformly distributed in $[0, 2\pi]$ and the
magnitude has PDF
\beq \label{eq:Loo_dist}
\begin{split}
f_{\mathrm{Loo}}(x) = \frac{8.686 x}{\Sigma_{\rm A} \sigma^2 \sqrt{2\pi}}& \int_0^{+\infty} \frac1{a} \ee^{-\frac{(20 \log_{10} (a) - M_{\rm A})^2}{2 \Sigma_{\rm A}^2} - \frac{x^2-a^2}{2 \sigma^2} }  \\
\times & I_0 \left( \frac{a x}{\sigma^2}\right) \dd a
\end{split}
\eeq where $M_{\rm A}$ and $\Sigma_{\rm A}$ are mean and standard
deviation of 
the Line-of-Sight (LoS) component, while
$\mathrm{MP} = 10 \log_{10}\sigma^2$ is the power of the scattered
component.  In \eqref{eq:Loo_dist}, $I_0(x)$ is the order-0 modified
Bessel function of the first kind.  The triple of parameters $(M_{\rm
  A}, \Sigma_{\rm A}, \mathrm{MP})$ characterizing the Loo
distribution is itself a random variable, with different PDF depending
on whether the state is good or bad, and sampled independently for
each state occurrence.  Correlation between successive channel
coefficients is obtained by low-pass filtering on the LoS component and Jakes' filtering on the scattered
component \cite{ITURRECP68111}. 

On the G2S link,  the signal received by satellite $\ell$ at time $t$ has SNR
\beq
\mathrm{SNR}_{\ell}(t) = \frac{P_{\rm GU} G_{\rm GU} G_{\ell} L_{\ell}(t) |h_{\ell}(t)|^2}{N_0 W_{\rm G2S}} = \rho_{\ell}(t) |h_{\ell}(t)|^2
\eeq
where
\begin{itemize}
\item $P_{\rm GU}$ is the GU TX power;
\item $G_{\rm GU}$ and $G_{\ell}$ are the antenna gains of the GU and of satellite $\ell$, respectively;
\item $L_{\ell}(t) = \left( \frac{c}{4\pi d_{\ell}(t)f_{\rm G2S}}\right)^2$ is the free-space path loss, with $c$ and $f_{\rm G2S}$ being the speed of light and the carrier frequency on the G2S link;
\item $h_{\ell}(t)$ is the term in the G2S channel that accounts for shadowing and fading;
\item $\rho_{\ell}(t)$ is the LoS SNR, without shadowing/fading;
\item $N_0$ is the noise power spectral density;
\item $W_{\rm G2S}$ is the signal bandwidth on the G2S channel.
\end{itemize}
 
\subsection{ISL Channel Model}
For the ISL channel, we adopt the model reported in
\cite{Akyildiz21}. In particular, we assume that the ISL works at a
large operating frequency, so that the depolarization effect can be
neglected. Moreover, the attenuation due to plasma frequency
and collision frequency is very low, compared to the free-space path
loss, thus it will also be neglected \cite{Akyildiz21}.   

The inter-satellite channel model boils down to an AWGN channel, with
random SNR due to pointing errors,  represented by a misalignment angle
$\xi$ between transmit and receive antennas.  We
assume that the antenna gain of both satellites on the ISL can be
characterized as \cite{Akyildiz21}
\begin{equation}
  G = G_0 \ee^{-\nu \xi^2}
\end{equation}
where $G_0$ is the maximum antenna gain (achieved when the main lobe
of the satellite antenna perfectly points towards the other
satellite), $\nu = 4 \ln 2 / \theta^2_{\rm 3dB}$ is a parameter related to the 3-dB width $\theta_{\rm 3dB}$ of the main
lobe  and $\xi$ is a Gaussian random variable with zero mean and
variance $\sigma^2_p$.

Thus, we can compute the received SNR on the ISL as
\beq \label{eq:snr_ISL}
\rho_{\rm ISL} = \frac{P_T L_{\mathrm{ISL}} G_0^2 \ee^{-2\nu \xi^2}
}{k_B T_0 W_{\mathrm{ISL}}}
\eeq
where
\begin{itemize}
\item $P_T$ is the transmit power  on the ISL link;
\item $L_{\mathrm{ISL}} = \left( \frac{c}{4\pi
  d_{\mathrm{ISL}}f_{\mathrm{ISL}}}\right)^2$ is the free-space path loss,  with
  $f_{\mathrm{ISL}}$ and $d_{\mathrm{ISL}} = 2(R_e+h_0) \sin
  \frac{\alpha_0}{2}$ being the carrier frequency and the link length
  on the ISL, respectively;
\item $k_B$ is the Boltzmann constant;
\item $T_0$ is the ambient noise temperature;
\item $W_{\mathrm{ISL}}$ is the signal bandwidth on the ISL.
\end{itemize}
We suppose  perfect receiver knowledge of the ISL SNR.

\subsection{Signal Received on the G2S Link}

In this subsection, we
describe the communication channel in the G2S link.

The GU takes its information word $\uv$ and encodes it with a
rate-$R_c$ channel code.  Then, the obtained codeword $\cv$ is modulated producing the sequence of symbols $x[n]$, $n=1,2,\ldots, N$, taken from a size-$S$ unit-energy modulation alphabet.
The $n$-th symbol is transmitted at time $t_n = nT_s$ where
$T_s$ is the symbol time.

In the soft-handover
scenario,  both satellites receive the signal transmitted
by the GU.  One of the two satellites works as relay (R), through the ISL,  according to the relaying strategies described in
Sect. \ref{sec:relay}; the other as destination (D).   More precisely,  when $\delta_{\rm S}(t) > \delta_{\rm T}(t)$, then $\rm R=T$ and $\rm D = S$.  Viceversa, when $\delta_{\rm S}(t) < \delta_{\rm T}(t)$, then $\rm R=S$ and $\rm D = T$.  Both satellites are supposed
to have perfect knowledge of their respective channel, because for
example it has been estimated through the use of pilots.  The
(normalized) signal received by satellite $\ell$ at time step $n$, $n = 1,\dots,N$ is given by
\begin{eqnarray}
  y_{\ell}[n] &=& \sqrt{\rho_\ell[n]} h_{\ell}[n] x[n] + w_\ell[n],
\end{eqnarray}
where $w_\ell[n]\sim \Cc\Nc(0,1)$, i.e., it is a circularly-symmetric complex Gaussian noise
sample with zero mean and unit variance,
$\rho_\ell[n]\triangleq\rho_{\ell}(nT_s)$ is the received LoS SNRs,
while $h_{\ell}[n]\triangleq h_\ell(nT_s)$ accounts for both shadowing
and fading. For the sake of notation compactness,  we define the vectors $\yv_\ell =
(y_\ell[1],\ldots,y_\ell[N])$,  with $\ell\in\{{\rm
  S},{\rm T}\}$ if we consider the physical satellites or $\ell\in\{{\rm
  R},{\rm D}\}$ if we refer to the logical satellites,  and analogously for $\rhov_\ell$ and
$\hv_\ell$.

\subsection{Relaying Techniques} \label{sec:relay}
With soft handover,  after reception from the GU,  the relaying satellite R may forward its received signal to
satellite D. We consider two possible choices for the relaying
strategy, as follows.
\begin{itemize}
\item \emph{Amplify and forward (AF)}: Satellite R works as a
  repeater, by only forwarding the received signal $y_{\rm R}[n]$.  The normalized (unit-energy) forwarded signal on the ISL link then
  becomes
  \beq \label{eq:x_isl}
  x_{\rm ISL}[n] = \frac{y_{\rm
      R}[n]}{\sqrt{q[n]}}, \quad\forall n 
      \eeq
  where $q[n] =
  \EE[|y_{\rm R}[n]|^2] = \rho_{\rm R}[n] |h_{\rm R}[n]|^2 + 1$ is a
  normalization factor.  Indeed,  with AF relaying, the normalized signal in \eqref{eq:x_isl} is amplified with an amplification gain before transmission.


\item \emph{Decode and forward (DF)}: Satellite R demodulates and
  decodes the received signal.  A block decoding error happens with
  probability $P_B$, according to the definition~\eqref{eq:PBe} given below.  We
  suppose that satellite R is aware whether decoding is successful or
  not.  If it is successful, it re-encodes the estimated
  information bits, modulates the coded bits and forwards the regenerated signal
  to S, so that
  \beq x_{\rm ISL}[n] = x[n], \quad \forall n
  \eeq
  If decoding is not successful, T forwards no signal to S.
\end{itemize} 

\subsection{Processing at Satellite D}
The (normalized) signal received by satellite D on the ISL from R at time step $n$ is
given by
\begin{equation}
  y_{\rm ISL}[n] = \sqrt{\rho_{\mathrm{ISL}}} x_{\rm ISL}[n] +
  w_{\rm ISL}[n],
\end{equation}
where $\rho_{\mathrm{ISL}}$ is defined in~\eqref{eq:snr_ISL}, and
$w_{\rm ISL}[n]\sim \Cc\Nc(0,1)$.  For DF,
$\rho_{\mathrm{ISL}}=0$ in the case of unsuccessful decoding at R.

In order to improve the link reliability, satellite D then performs
maximum ratio combining (MRC) of the two signals, the one received
directly from the GU on the G2S link and the one relayed by R on the
ISL link. For both AF and DF, after suitable normalization, we can
write the MRC output as
\begin{equation} \label{eq:mrc}
\widetilde{y}_{\vartheta}[n] = \widetilde{h}_{\vartheta}[n] x[n] + \widetilde{w}[n],
\end{equation}
where $\vartheta \in \{{\rm AF}, {\rm DF}\}$, $\widetilde{w}[n]$ is a zero-mean complex Gaussian noise with variance 1
and $\widetilde{h}[n]$ is the equivalent channel at time step $n$.  In the case of AF, we have
\beq
\widetilde{h}_{\rm AF}[n] = \sqrt{\rho_{\rm D}[n] |h_{\rm D}[n]|^2 + \frac{\rho_{\mathrm{ISL}} \rho_{\rm R}[n] |h_{\rm R}[n]|^2}{\rho_{\mathrm{ISL}} + q[n]}}
\eeq
\begin{equation}
\begin{split}
\widetilde{y}_{\rm AF}[n] = &\sqrt{\rho_{\rm D}[n] }\left( \frac{h_{\rm D}[n]}{\widetilde{h}_{\rm AF}[n]}\right)^*  y_{\rm D}[n] + \\ &+\frac{\sqrt{\rho_{\mathrm{ISL}} q[n]\rho_{\rm R}[n]}  }{\rho_{\mathrm{ISL}} + q[n]} \left(\frac{h_{\rm R}[n]}{\widetilde{h}_{\rm AF}[n]} \right)^* y_{\rm ISL}[n].
\end{split}
\end{equation}
Instead, for DF,
\beq
\widetilde{h}_{\rm DF}[n] = \sqrt{\rho_{\rm D}[n] |h_{\rm D}[n]|^2 + \rho_{\mathrm{ISL}} }
\eeq  
\begin{equation}
\widetilde{y}_{\rm DF}[n] = \sqrt{\rho_{\rm D}[n] }\left( \frac{h_{\rm D}[n]}{\widetilde{h}_{\rm DF}[n]}\right)^* y_{\rm D}[n] + \frac{\sqrt{\rho^{(\mathrm{ISL})}}}{\widetilde{h}_{\rm DF}[n]} y_{\rm ISL}[n].
\end{equation}
Notice that AF requires the knowledge of the channel $h_{\rm R}[n]$ at
satellite D, while DF does not.  

\section{BLER analysis}

In this section, we derive the final performance in terms of BLER for the combined
signal at satellite D. Consider the AWGN channel $\Yc = \sqrt{\rho}\Hc \Xc + \Wc$ with $\Wc \sim \Cc\Nc(0,1)$ and $\Xc$ belonging to a size-$S$ modulation $\Sc$. Let $H \left(\Xc |  y,  \sqrt{\rho} h  \right)$ be the equivocation on such channel when $\Yc = y$ and $\Hc = h$, i.e.,  
\beq
H \left(\Xc |  y,  \sqrt{\rho} h  \right) = -\sum_{s \in \Sc} p\left(s | y,  \sqrt{\rho} h\right) \log_2 p\left(s | y,  \sqrt{\rho}h\right)
\eeq
where
\beq
p\left(s | y,  \sqrt{\rho} h\right) = \frac{\ee^{-| y - \sqrt{\rho} h s|^2}}{\sum_{s' \in \Sc} \ee^{-| y - \sqrt{\rho} h s'|^2}}
\eeq
Then, the (conditional) mutual information (MI) per bit is given by
\beq \label{eq:capcond1}
C(y,  \sqrt{\rho} h) = 1 - \frac1{\log_2 S}  H \left(\Xc |  y,  \sqrt{\rho} h \right)
\eeq

Suppose a channel code of length $N_c$ symbols is transmitted on the AWGN channel and denote $\yv, \hv, \rhov$  the length-$N_c$ vectors of realized received samples, channel coefficients and SNR values, respectively. We assume that the channel decoder has a block error probability given by\footnote{In \eqref{eq:PBe}-\eqref{eq:capcond}, we have used the Hadamard (elementwise) product: for two equal-length vectors $\av$ and $\bv$,  $\av \odot \bv = \cv$ where $\cv$ is another vector with $i$-th entry $c_i = a_i b_i$.} 
\beq \label{eq:PBe} P_B(e| \yv, \sqrt{\rhov}\odot \hv ) =
\left\{ \begin{array}{ll} 1, & \overline{C}\left(\yv, \sqrt{\rhov}\odot \hv
  \right) \leq C_{T} \\ 0, & \overline{C}\left(\yv, \sqrt{\rhov}\odot \hv \right)
  > C_{T}
\end{array}\right.
\eeq
where 
\beq \label{eq:capcond}
\overline{C}(\yv, \sqrt{\rhov}\odot \hv) = \frac1{N_c} \sum_{n=1}^{N_c} C(y[n],  \sqrt{\rho[n]}h[n] )
\eeq
gives the average MI per bit on the code block. In \eqref{eq:PBe}, $C_T$ is  the threshold MI on the AWGN channel for the channel code. Intuitively, \eqref{eq:PBe} means that if the realized flow of information on the transmission channel is large enough, the signal is correctly decoded, otherwise it is lost. Typically, \eqref{eq:PBe} is a good approximation of the real behavior of codes with large block length, which is of interest in NTNs.     

Owing to \eqref{eq:mrc}, the performance in terms of BLER of the soft-handover scheme will be given by
\beq
P_{B,  \mathrm{SH}, \vartheta}(e ) = \EE_{\widetilde{\yv}_{\vartheta}, \widetilde{\hv}_{\vartheta}} P_B(e | \widetilde{\yv}_{\vartheta}, \widetilde{\hv}_{\vartheta})
\eeq
where $\vartheta \in \{{\rm AF}, {\rm DF}\}$ and  $\EE_{z}$ is the average with respect to random variable $z$.  

As a comparison,  the BLER performance for hard handover is given by
\beq
P_{B, \mathrm{HH}}(e ) = \EE_{\yv_{\rm D}, \hv_{\rm D}, \rhov_{\rm D}} P_B(e | \yv_{\rm D}, \sqrt{\rhov_{\rm D}} \odot \hv_{\rm D} ).
\eeq

\subsection{Soft Handover  with Noiseless ISL \label{sec:theory}}

In this subsection, we derive the performance of soft handover, for both AF and DF, in the limit of $\rho_{\mathrm{ISL}} \rightarrow \infty$.  In such case, the considered system is equivalent to a virtual 
$1 \times 2$ single-input multiple-output system, in which there is only one satellite with two antennas,  and the received signal is $(\yv_D,\yv_R)$.  Notice that, given the elevations of the two satellites and thus the values of $(\rhov_D,\rhov_R)$, the two branches of the received signal are independent in our model. 

In such a scenario, the optimal receiver performs MRC before decoding, as in the AF case.   The instantaneous SNR after MRC will then be given by
\beq
\lim_{\rho_{\mathrm{ISL}} \rightarrow \infty} |\widetilde{h}_{\rm AF}[n]|^2 = \rho_{\rm D}[n] |h_{\rm D}[n]|^2 +\rho_{\rm R}[n] |h_{\rm R}[n]|^2
\eeq 
The advantage of AF-based soft handover  with respect to hard handover, on a noiseless ISL,   is then obvious,  as the hard-handover SNR is equal to $\rho_{\rm D}[n] |h_{\rm D}[n]|^2$, so that hard handover is more prone to B-state channel conditions on the link between GU and satellite D.  Notice that this conclusion can be generalized to a noisy ISL,  as the SNR after MRC will always be larger than in the  hard-handover case.   

On the noiseless ISL,  the DF strategy corresponds to separate decoding on $\yv_{\rm D}$ and $\yv_{\rm R}$, so that the transmitted block is lost if and only if both decoding attempts fail.  
 Thus:
\beq
\begin{split}
\lim_{\rho_{\mathrm{ISL}} \rightarrow \infty} P_{B,  \mathrm{SH}, {\rm DF}}(e ) = &\EE_{\rhov_{\rm D}, \rhov_{\rm R}} \left\{ \EE_{\yv_{\rm D}, \hv_{\rm D}} P_B(e| \yv_{\rm D}, \sqrt{\rhov_{\rm D}} \odot \hv_{\rm D}) \right. \\ 
&\left. \times \EE_{\yv_{\rm R}, \hv_{\rm R}}P_B(e| \yv_{\rm R}, \sqrt{\rhov_{\rm R}} \odot \hv_{\rm R}) \right\}
\end{split}
\eeq

While DF proves suboptimal with respect to AF for a noiseless ISL,  on a noisy ISL this may not be true in general, as the latter combines the noise on the ISL with the noise on the G2S link between the GU and satellite R.

\section{Simulation results\label{sec:results}}

In Table \ref{tab:simulation_parameters},  we report the main parameters of the simulation.

We consider a single satellite pass\footnote{We chose a satellite pass with favorable conditions.  Considering the definitions of Section \ref{sec:geo_model}, and the parameter values of Table \ref{tab:simulation_parameters},  this pass happens for $18000 \,{\rm s} <t < 19000 \,{\rm s}$.}.  In that pass, we focus on the
time window for which satellites T and S have the highest elevations (both larger than $\delta_{\min}$) in the constellation.  The elevation
time series for both satellites is quantized on the set of values
$\{30^{\circ},45^{\circ},60^{\circ},70^{\circ}\}$, for
which there exist tables of channel parameters in
\cite{ITURRECP68111}.  More precisely, we consider a carrier frequency
$f_{\rm G2S} = 2.2$\,GHz,  and the suburban environment, so that the parameter
tables for the different elevations can be found in \cite[Annex 2,
  Sect. 2.2]{ITURRECP68111}.  Given the parameters, the channel
coefficients for the two G2S links are then generated independently.
The visibility interval of both satellites lasts for about 2 minutes, during which we suppose that 500 code blocks are transmitted. 

In the simulations, we consider as the independent variable a
reference SNR, which is the LoS SNR when the satellite is at the
minimum distance (i.e.,  the maximum elevation) from the GU.  In other
words, the reference SNR for satellite $\ell=\{{\rm S}, {\rm T}\}$ is
given by
\begin{equation}
  \rho_{\mathrm{ref}} = \frac{P_{\rm GU} G_{\rm GU}
    G_{\ell} L_{\mathrm{ref}}}{N_0 W_{\rm G2S}}
\end{equation}
where
\begin{equation}
L_{\mathrm{ref}} = \left( \frac{c}{4\pi h_0 f_{\rm G2S}}\right)^2
\end{equation}
so that the actual LoS SNR is $\rho_\ell[n] = \rho_{\mathrm{ref}}
L_{\ell}[n] / L_{\mathrm{ref}}$.  With the parameters reported in Table \ref{tab:simulation_parameters},  a reference SNR of 20 dB corresponds to a transmitted power $P_{\rm GU}$ of about 50 mW. 

\begin{table}[th]
\centering
\caption{Simulation parameters.}
\begin{tabular}{|l|c|c|}
\hline
Parameter & Notation  & Value \\
\hline \hline
\multicolumn{3}{|c|}{Geometrical model} \\
\hline
Orbit height & $h_0$ &  550\,km\\
Orbit inclination & $i_{\rm K}$ & $45^{\circ}$ \\
Constellation size & $M$ &  $\{ 42, 63\}$ \\
GU coordinates & $(\lambda_{\rm GU},\phi_{\rm GU})$ & ($45^{\circ}$ N,  $7^{\circ}$  E) \\
Minimum elevation & $\delta_{\min} $ & $25^{\circ}$ \\
\hline
\multicolumn{3}{|c|}{Ground-to-satellite (G2S) link} \\
\hline
Carrier frequency & $f_{\rm G2S}$ & 2.2\,GHz\\
Transmitted power & $P_{\rm GU}$  & See text \\
GU antenna gain & $G_{\rm GU}$ & 0 dBi \\
Satellite antenna gain & $G_{\rm S},  G_{\rm T}$ & 50 dBi \\
Signal bandwidth & $W_{\rm G2S}$ & 5\,MHz \\
Environment & - & Suburban \\
Channel model parameters & - & From \cite{ITURRECP68111} \\
Noise level & $N_0$ & -174\,dBm/Hz\\
\hline
\multicolumn{3}{|c|}{Inter-satellite link (ISL)} \\
\hline
Carrier frequency & $f_{\rm ISL}$ & $\{2,193\}$ THz \\
Transmitted power & $P_{T}$  & $\{5, 10 , 15,20, 25\}$ dBW \\
Ambient noise temperature & $T_0$ & 7000\,K \\
Signal bandwidth & $W_{\mathrm{ISL}}$ & 2\% $f_{\rm ISL}$  \\
TX/RX antenna gain & $G_0$ & $\{60, 90\}$ dBi\\
Half-power beam width & $\theta_{\rm 3dB}$ &  $202.5 \times 10^{-G_0/20}$ deg\\
Misalignment variance & $\sigma_p^2$ & See text \\
\hline
\multicolumn{3}{|c|}{Transmitted signal format} \\
\hline
Channel coding rate & $R_c$ & 1/2 \\
Channel code & - & (3,6) LDPC code \\
Code threshold & $C_T$ & 0.5714\,bits \\ 
Modulation format & - & QPSK ($S = 4$) \\
\hline
\end{tabular}
\label{tab:simulation_parameters}
\end{table}

\subsection{Comparison between AF and DF \label{sec:AFvsDF}} 

We first compare AF and DF in the scenario summarized by the parameters in Table \ref{tab:simulation_parameters}. In particular, we consider the case in which: 
\begin{itemize}
\item there is no misalignment in the ISL; 
\item the angular distance between the two satellites is $\alpha_0 = 0.15$ rad, corresponding to (about) $M = 42$ satellites in the same orbit;
\item the carrier frequency on the ISL is $f_{\mathrm{ISL}} = 193$ THz (in the optical band), with an antenna gain $G_0 = 90$ dB.
\end{itemize}   

\begin{figure}[t]
  \centering
\includegraphics[width=1.01\columnwidth]{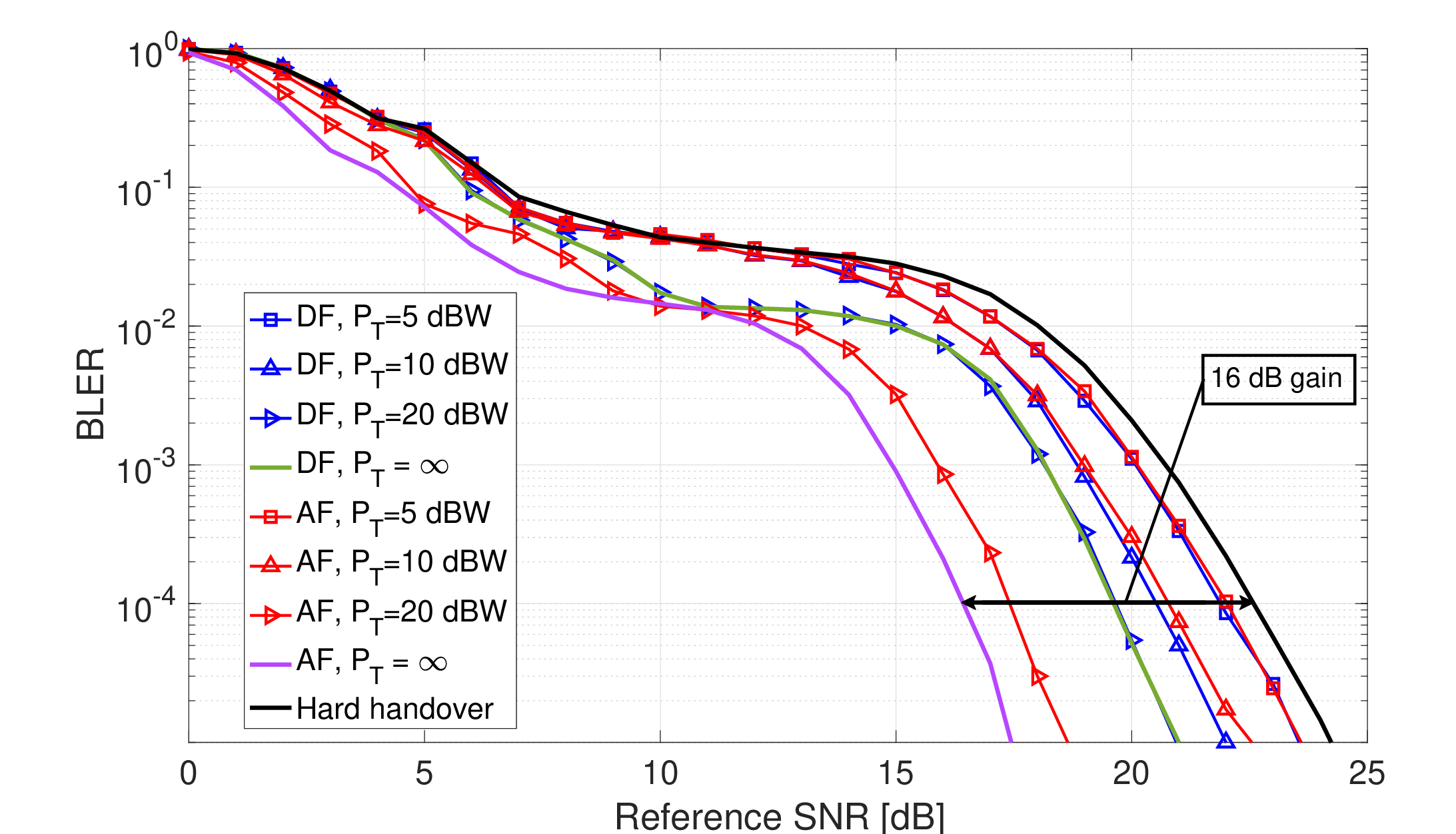}
\caption{Comparison between AF and DF with $f_c = 193$ THz and $N = 42$ satellites.}
\label{fig:AF_vs_DF}
 \end{figure}

Figure \ref{fig:AF_vs_DF} shows the numerical results. Blue curves represent the BLER for DF, with three different values of the transmit power $P_{T}$ on the ISL, namely $P_{T} \in \{5, 10, 20\}$ dBW. For the same values of $P_{T}$, red curves depict the performance of AF. The thick black curve shows the BLER of hard handover, while the thick purple and green curves correspond to the BLER for AF and DF,  respectively,  with an infinite SNR on the ISL (or, equivalently, a noiseless ISL).  

As it can be seen, at $\mathrm{BLER} = 10^{-4}$ and with the parameter setting of the figure, the ultimate gain of  soft handover, compared to hard handover, is 16 dB.  Most of this potential advantage can be achieved by adopting an AF relaying strategy with a transmitted power of $P_{T} = 20$ dBW. The comparison between red and blue curves reveals that,  for low-to-intermediate transmitted power ($P_{T} = 10$ dBW),  AF and DF show a comparable performance.  Instead,  for large $P_{T}$,  AF improves on DF,  as the former seems to exploit better the diversity gain offered by multisatellite reception, in accordance with the analysis of Section \ref{sec:theory}.  

It is worth commenting the shape of the curves for different SNR values.  For reference SNR lower than 10 dB,  BLER curves decrease with SNR as expected,  but for middle SNR values (say,  around 10 to 12-15 dB, depending on the curve) there is a floor.  
This is due to the fact that,  in the elevation time series,  there is a time window  where both satellites have a quantized elevation as low as $45^{\circ}$,  so that the channel model for both satellites shows a more severe impact of B-state events.  As a result,  the error floor appears when the reference SNR is not  large enough to yield a good performance also in this time window.  

 \subsection{Increasing the Size of the Constellation \label{sec:large_const}}
 
 \begin{figure}[t]
  \centering
\includegraphics[width=1.01\columnwidth]{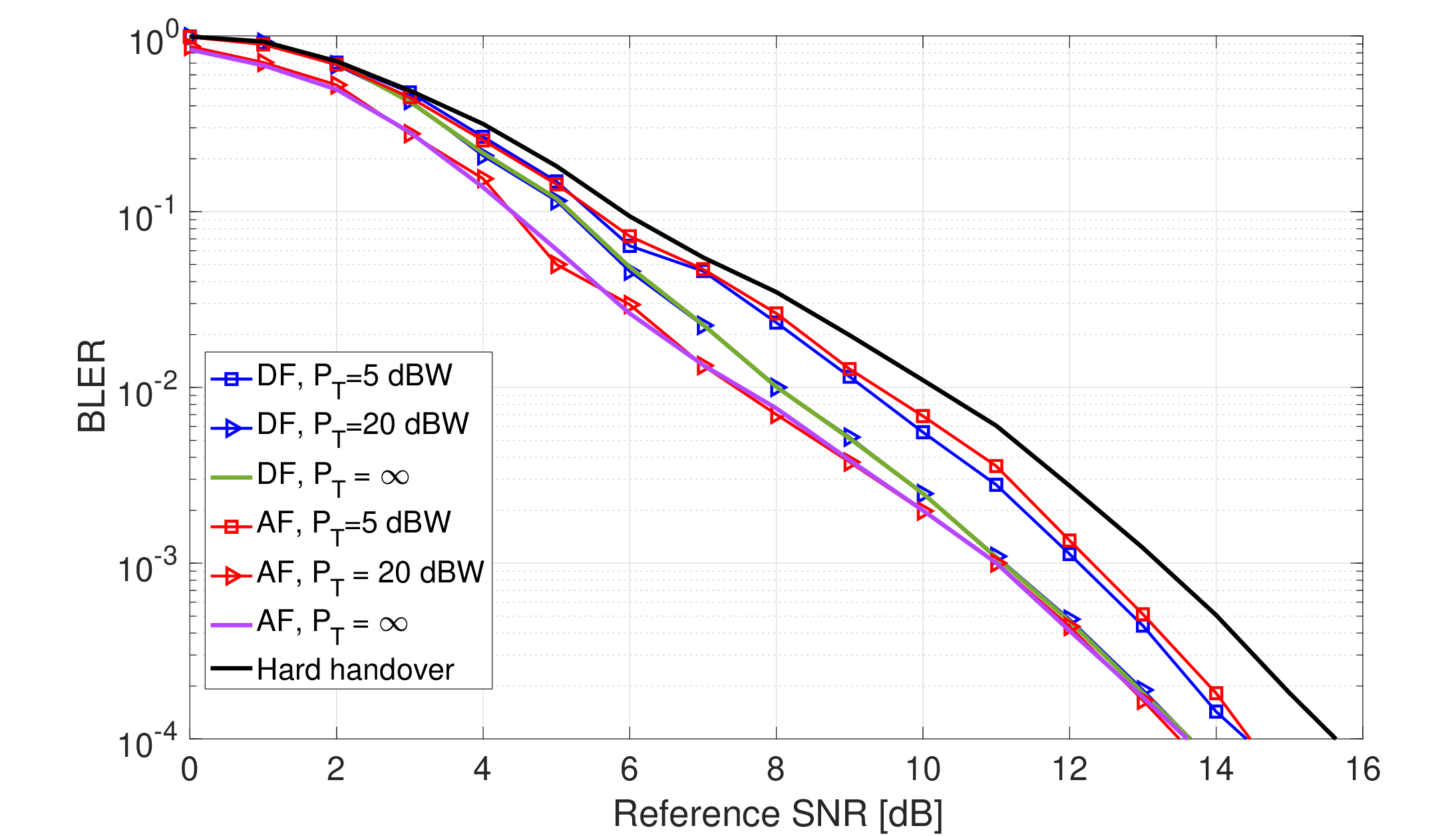}
\caption{Comparison between AF and DF with $f_c = 193$ THz and $M = 63$ satellites.}
\label{fig:AF_vs_DF_63}
 \end{figure}

 After obtaining in the previous subsection the BLER performance for a  number of satellites in the same orbit equal to $M = 42$, in this subsection we obtain the same results for $M = 63$, corresponding to an angular distance between neighbor satellites of $\alpha_0 = 0.1$ rad. In both cases, the elevation time series for the two satellites are very similar, with one delayed with respect to the other. For $M = 63$, when the two satellites have about the same elevation, this elevation is close to $60^{\circ}$, while for $M= 42$ it is close to $45^{\circ}$,  as already mentioned in Section \ref{sec:AFvsDF}.  All the other parameters are as in Figure \ref{fig:AF_vs_DF}.

 Figure \ref{fig:AF_vs_DF_63} shows the BLER performance in the new scenario with more satellites. Blue curves are obtained for DF while red curves are for AF. As it can be seen, the maximum achievable gain of soft handover is reduced to about 2--2.5 dB.   In this scenario,  DF slightly outperforms AF for $P_{\rm T} = 5$ dBW,  as its performance is not affected by the noise on the GU-to-T link, while for larger $P_{\rm T}$ AF is better than DF,  as in Figure \ref{fig:AF_vs_DF}.  Overall,  we can say that increasing the size of the constellation makes soft handover less awarding, as, for $N=63$ satellites, a BLER of $10^{-3}$ is obtained with hard handover for a reference SNR equal to about 13  dB,  while the same value for $N=42$ satellites requires a reference SNR of at least 15 dB and a large ISL transmitted power (with AF).  Also, we conclude that,  in the simulated scenario,  DF does not achieve the performance gain that would repay for its complexity increase with respect to AF.
 
Finally notice that, in this scenario, since there is always at least one satellite with quantized elevation larger than $45^{\circ}$, there is no error floor for middle SNR values.

\subsection{The Impact of Misalignment} 
 
 \begin{figure}[t]
  \centering
\includegraphics[width=1.01\columnwidth]{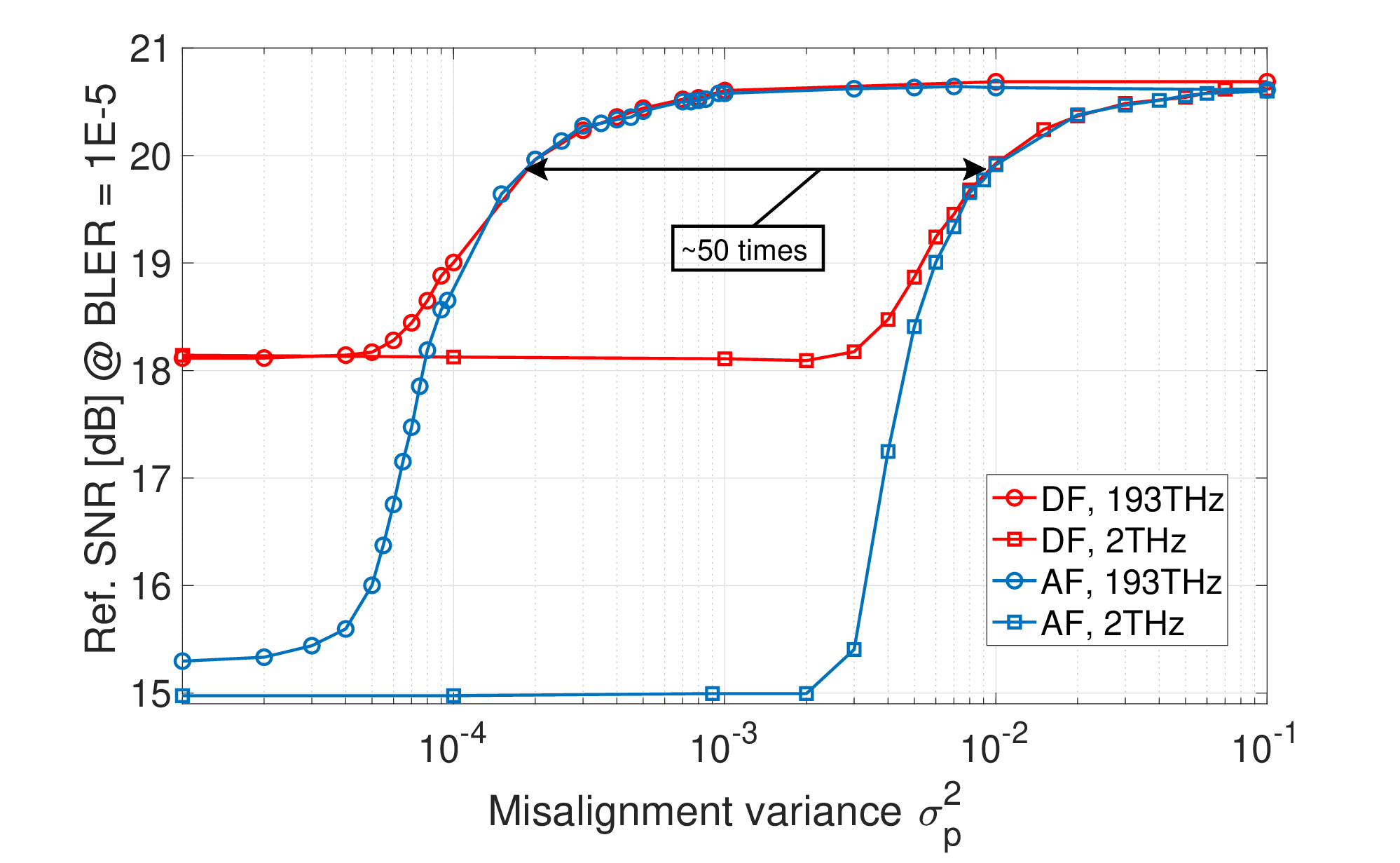}
\caption{The scenario with misalignment on the ISL. Parameters: $M = 42$, $P_T = 25$ dBW.}
\label{fig:misal_42}
 \end{figure} 
 
 \begin{figure}[t]
  \centering
\includegraphics[width=1.01\columnwidth]{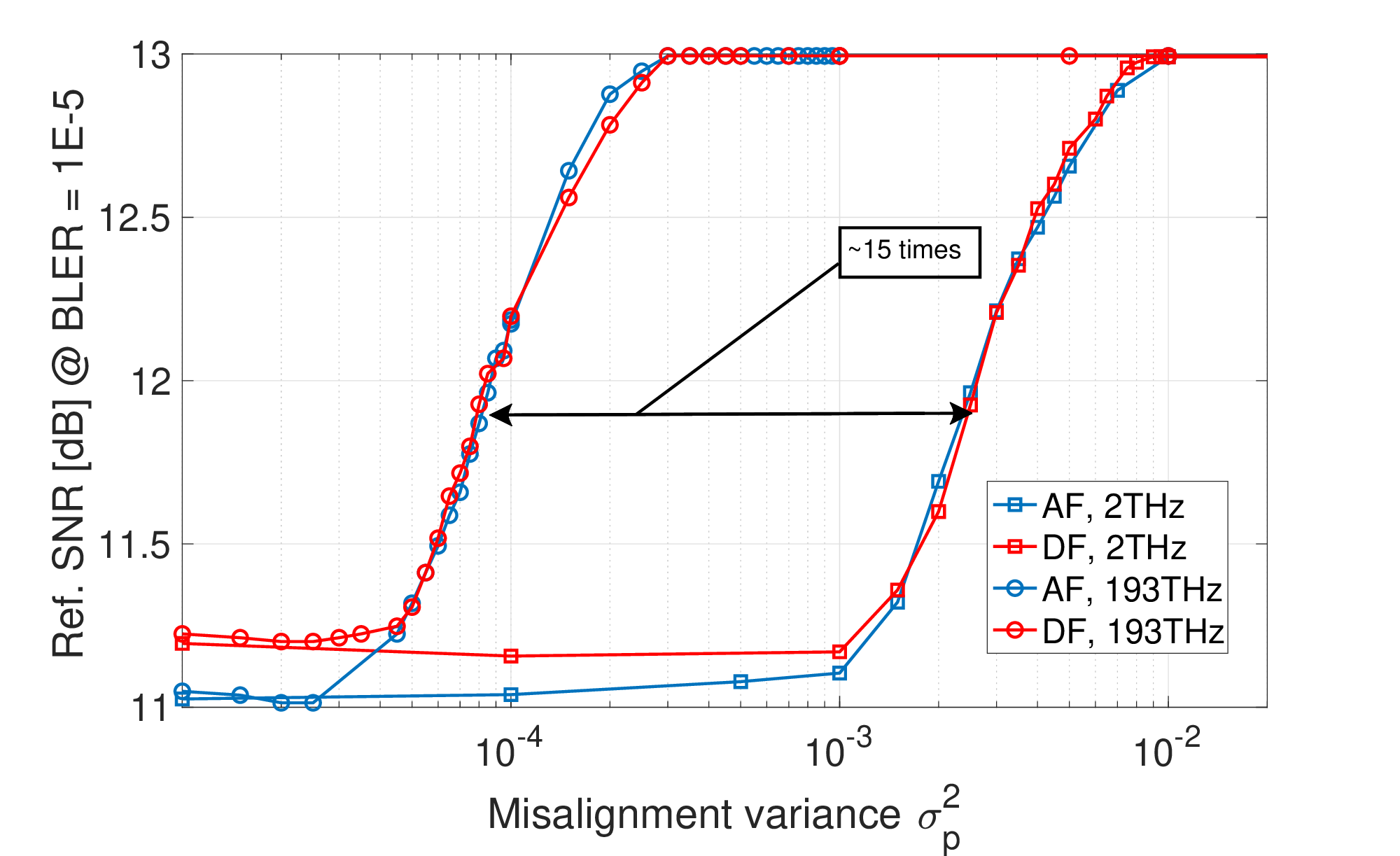}
\caption{The scenario with misalignment on the ISL.  Parameters: $M = 63$, $P_T = 15$ dBW.}
\label{fig:misal_63}
 \end{figure}

 Finally, we consider the impact on  performance of misalignment in the ISL.  Following \cite{Akyildiz21}, we consider two different carrier frequencies on the ISL, namely, $f_{\mathrm{ISL}} = 193$ THz, as in Sections \ref{sec:AFvsDF} and \ref{sec:large_const}, and $f_{\mathrm{ISL}} = 2$ THz.  In both cases,  we adopt Kraus' approximate formula \cite{Kraus-2002-AntennasB} to derive the half-power beamwidth given the antenna gain, as shown in Table~\ref{tab:simulation_parameters}.   For $f_{\mathrm{ISL}} = 193$ THz,  a gain of $90$ dB yields a half-power beamwidth of $0.0064^{\circ}$,  while for $f_{\mathrm{ISL}} = 2$ THz,  we have set a gain $G_0 = 60$ dB,  which results in a half-power beamwidth equal to $0.2^{\circ}$.   The misalignment stays constant over a single code block.    
 
For the two bands,  Figures \ref{fig:misal_42}-\ref{fig:misal_63} show, for both DF and AF,  the reference SNR needed to achieve $\mathrm{BLER} = 10^{-5}$,  as a function of $\sigma_p^2$,  the misalignment variance. In Figure \ref{fig:misal_42}, which is for $M = 42$ satellites, the transmitted power on the ISL is set to $P_T = 25$ dBW, while in Figure \ref{fig:misal_63}, for $M = 63$ satellites, $P_T = 15$ dBW.  The figures show similar behavior in the two bands.  For low enough $\sigma_p^2$,  the performance reaches that of the no-misalignment case. Instead, for sufficiently high  $\sigma_p^2$, the soft-handover performance reaches the hard-handover BLER,  with no help anymore from the ISL.  As it was to be expected,  the optical band is much more sensitive,  due to the narrower beam. For instance, with the parameter settings of Figure \ref{fig:misal_63} and threshold reference SNR equal to 12 dB, a 15 times larger misalignment variance can be tolerated in the case with $f_{\mathrm{ISL}} = 2$ THz, compared to the case with $f_{\mathrm{ISL}} = 193$ THz.  It is worth noting that, in both figures and  for $\sigma_p^2 = 10^{-3}$,  the performance in the THz band is still optimal,  while the optical scenario shows already a completely spoiled contribution of the ISL.  Of course,  the larger bandwidth in the optical band allows to serve more users at the same time,  and to potentially reduce the number of satellites connected to the ground by  feeder links.  

It is also worth noting that, when misalignment starts to worsen performance, the difference between AF and DF disappears.

\section{Conclusions\label{sec:conclusion}}

We considered a constellation of LEO satellites connected to a handheld device on the ground.  In particular,  we studied the benefits of soft handover in the UL from the physical-layer point of view.  We have performed simulation results in a realistic scenario, where the impairments are taken into account both in  the ground-to-satellite and the ISL.  

Our results show that, although implying a larger
computation complexity, soft handover can achieve a substantial
performance gain  over hard handover.  Considering different relaying techniques,  AF should be preferred to DF,  as the former shows essentially the same performance of the latter for low-to-medium transmitted power on the ISL,  and a definite advantage for large transmitted power.  Moreover,  AF has a lower complexity for the relaying satellite.  Adding satellites to the constellation improves the performance of both hard handover and soft handover,  and reduces the achievable gain of soft handover.  Finally,  our results show how important is to maintain a suitable alignment of the satellite antennas, to achieve the potential gain of soft handover, especially in the optical band.


\section*{Acknowledgment}

Funded by the European Union, under ANTERRA 101072363 HORIZON-MSCA-2021-DN-01. Views and opinions expressed are however those of the authors only and do not necessarily reflect those of the European Union. Neither the European Union nor the granting authority can be held responsible for them.

\bibliographystyle{IEEEtran}
\bibliography{sh_leo}

\end{document}